\documentclass[aps,showpacs]{revtex4}

\usepackage{amsmath,amssymb}
\usepackage{tabularx}
\usepackage{epsfig}
\usepackage{graphicx}

\begin{document}

\title{A brief note on how to unify dark matter, dark energy and inflation}

\author{Grigoris Panotopoulos\footnote{grigoris@theorie.physik.uni-muenchen.de}}

\date{\today}

%\address{~}

\address{ASC, Department of Physics LMU, Theresienstr. 37, 80333 Munich, Germany}

\begin{abstract}
A scenario in which inflation, dark energy and dark matter can be unified into a single scalar field, the inflaton field $\phi$, is studied. The inflaton is identified with the sneutrino, the scalar partner of the heavy neutrino. We determine the conditions needed for avoiding the gravitino problem and not having negligible plasma effects and we obtain the allowed range for the sneutrino coupling.
\end{abstract}


\maketitle

One of the theoretical problems in modern cosmology is to understand the nature of cold dark matter in the universe. There are good reasons, both observational and theoretical, to suspect that a fraction of $0.22$ of the energy density in the universe is in some unknown ``dark'' form. Furthermore, it is by now an established observational fact that the present universe undergoes an
accelerating phase. During the last ten years or so a remarkable progress in technology has allowed us to witness extraordinary
precision measurements in cosmology. A plethora of observational data are now available~\cite{data}, which show that we live in a
flat universe that expands with an accelerating rate and that the dominant component in the energy budget of the universe
is an unusual material, the nature of which still remains unknown. Identifying the origin and nature of dark energy is
one of the great challenges in modern theoretical cosmology.

Inflation~\cite{inflation} has become the standard paradigm for the early universe, because it solves some outstanding problems present in the standard Hot Big-Bang cosmology, like the flatness and horizon problems, the problem of unwanted relics, such as magnetic monopoles, and produces the cosmological fluctuations for the formation of the structure that we observe today. The recent spectacular CMB data from the WMAP satellite have strengthen the inflationary idea, since the observations indicate an \emph{almost} scale-free spectrum of Gaussian adiabatic density fluctuations, just as predicted by simple models of inflation. However, a yet unsolved problem about inflation is that we do not know how to integrate it with ideas in particle physics. For example, we would like to identify the inflaton, the scalar field that drives inflation, with one of the known fields of particle physics.

Usually we invoke different models to explain dark matter, dark energy and inflation. So we attribute cold dark matter to a supersymmetric particle which is color and charge neutral (neutralino~\cite{neutralino}, axino~\cite{axino}, gravitino~\cite{gravitino}), inflation is driven by a scalar field (inflaton) which rolls down its potential and for dark energy another scalar field (k-essence~\cite{kessence} or tachyon~\cite{tachyon} or quintessence~\cite{quintessence}) is responsible. However it is advantageous to have a model to explain more than one things simultaneously. For example there are efforts to unify inflation and dark energy in quintessential inflation~\cite{quininfl} or to unify dark matter and dark energy using Chaplygin gas~\cite{cg} or quintessino~\cite{quintessino}, the superpartner of quintessence.

In a recent work~\cite{liddle} the conditions needed to unify the descriptions of inflation, dark energy and dark matter were considered. The conclusion of that work is that the main requirement for a working scenario is an incomplete inflaton decay. The problem however is that using the usual preheating or reheating theory one obtains results that contradict observations~\cite{liddle, cardenas}. Very recently it was pointed out~\cite{cardenas} that there is still the possibility of having a unifying description of dark matter, inflation and dark energy, provided that the effect of plasma masses is taken into account. In this brief note we propose that the sneutrino plays the role of the inflaton field, we determine the conditions needed for avoiding the gravitino problem and not having negligible plasma effects and finally we obtain the allowed range for the sneutrino coupling.

According to chaotic inflation with a potential for the inflaton field $\phi$ of the form $V=(1/2) M^2 \phi^2$, the WMAP normalization condition requires for the inflaton mass that $M \sim 10^{13}$~GeV. One of the most exciting  experimental results in the last years has been the discovery of neutrino oscillations~\cite{Fukuda:1998mi}. These results are nicely explained if neutrinos have a small but finite mass~\cite{Ahmad:2002jz}. The simplest models of neutrino masses invoke heavy gauge-singlet neutrinos that give masses to the light neutrinos via the seesaw mechanism~\cite{seesaw}. If we require that light neutrino masses $\sim 10^{-1}$ to $10^{-3}$~eV, as indicated by the neutrino oscillations data, we find that the heavy singlet neutrinos weigh $\sim 10^{10}$ to $10^{15}$~GeV, a range that includes the value of the inflaton mass compatible with WMAP. On the other hand, the hierarchy problem of particle physics is elegantly solved by supersymmetry (see e.g.~\cite{Martin:1997ns}), according to which every known particle comes with its superpartner, the sparticle. In supersymmetric models the heavy singlet neutrinos have scalar partners with similar masses, the sneutrinos, whose properties are ideal for playing the role of the inflaton~\cite{Murayama:1992ua}.

Let us collect now the formulas that we shall be using throughout our discussion. Assuming that the sneutrino decays into leptons and Higgs particles, $\tilde{N} \rightarrow l~H$, through Yukawa couplings, then the decay rate is given by the well-known result (in the zero temperature limit)
\begin{equation}
\Gamma_0 = \frac{f^2 M}{4 \pi} \equiv \alpha M
\end{equation}
where $M \sim 10^{13}$~GeV is the sneutrino inflaton mass, $\alpha=f^2/(4 \pi)$ and $f < 1$ is the Yukawa coupling in perturbation theory. When plasma effects are taken into account the decay products acquire an effective mass $m(T) \sim g T$~\cite{hotplasma}, with $g$ a typical  gauge coupling constant and $T$ the plasma temperature, and the decay rate is then given by~\cite{riotto}
\begin{equation}
\Gamma(T) = \Gamma_0 \sqrt{1-4 \frac{m(T)^2}{M^2}}
\end{equation}
We see that the decay is kinematically allowed as long as $M > 2 m(T)$ and that the effect is negligible in the limit $M \gg m(T)$. Apart from the inflaton mass two more quantities are of central importance in our presentation, namely the reheating temperature after inflation $T_R$ and the maximum temperature during reheating $T_{max}$. These are given by~\cite{riotto, giudice}
\begin{eqnarray}
T_R & = & 0.2 \left ( \frac{100}{g_{eff}} \right )^{1/4} \alpha^{1/2} \sqrt{M m_{pl}} \\
T_{max} & = & 0.6 \alpha^{1/4} \left ( \frac{m_{pl} V^{1/2}}{g_{eff} M^3} \right )^{1/4} M
\end{eqnarray}
where $m_{pl}=1.2 \times 10^{19}$~GeV is Planck mass, $g_{eff}$ is the number of relativistic degrees of freedom and $V$ is the value of the inflaton potential at the end of inflation. It is easy to check that for $f < 1$ it is always $T_{max} > T_R$. According to the values of $M, T_R, T_{max}$ there are three possibilities for their hierarchy, $T_{max} < M$ (case I), $T_R < M < T_{max}$ (case II) and $M < T_R$ (case III). Furthermore, a viable inflationary model should avoid the gravitino problem~\cite{Khlopov:1984pf}. The gravitino constraint imposes an upper bound on $T_R$ or $M$ depending on the case. In particular, for cases I,II one obtains a bound on $T_R$, $T_R \leq (10^8-10^9)$~GeV~\cite{riotto} while for case III one obtains a bound on $M$, $M \leq (10^8-10^9)$~GeV~\cite{riotto}.

Now we are able to determine the conditions needed for in incomplete reheating compatible with the gravitino constraint. In case (I) the effect of the plasma effect is negligible. In case (III) the WMAP normalization $M \sim 10^{13}$~GeV  contradicts the gravitino constraint $M \leq (10^8-10^9)$~GeV. Finally, in case (II) the requirement $T_{max} > M$ yields for the inflaton coupling $\alpha > 1.6 \times 10^{-8}$, while the gravitino constraint $T_R \leq (10^8-10^9)$~GeV yields $\alpha \leq 2 \times 10^{-15}$. Obviously the two bounds just obtained are incompatible with each other and thus it seems that it is impossible to satisfy all requirements simultaneously. However there is a way out if we assume that the axino is the lightest supersymmetric particle (LSP) and that the gravitino is the next-to-lightest supersymmetric particle. The axino is the superpartner of the axion~\cite{wilczek} which solves the QCD problem via the Peccei-Quinn mechanism~\cite{quinn}. In this case it has been shown~\cite{Asaka:2000ew} that although the gravitino decays to axion and axino with a lifetime larger than the primordial nucleosynthesis time, it does not destroy any BBN light nuclei because the axion and the axino are harmless due to their super-weak coupling to ordinary particles. Furthermore un upper bound on the reheating temperature was obtained, $T_R \leq 10^{15}$~GeV~\cite{Asaka:2000ew}, which is much weaker than that obtained in the minimal supersymmetric standard model. This upper bound is required in order not to have $^4 \textrm{He}$ overproduction. Now if we adopt this assumption for axino being the LSP, in case (II) it is sufficient to require $T_R < M$, which yields for the inflaton coupling $\alpha < 5 \times 10^{-5}$. Therefore we obtain for the inflaton coupling $\alpha$ the allowed range
\begin{equation}
1.6 \times 10^{-8} < \alpha < 5 \times 10^{-5}
\end{equation}
or correspondingly for the Yukawa coupling $f=\sqrt{4 \pi \alpha}$ we obtain the range
\begin{equation}
4.5 \times 10^{-4} < f < 2.5 \times 10^{-2}
\end{equation}
To summarize, in this brief note we have elaborated on the conditions needed for an incomplete reheating and for avoiding the gravitino problem and we have proposed the scalar neutrino as an natural candidate for the inflaton field. Assuming that the axino is the LSP and that the gravitino is the NLSP we have been able to satisfy all the requirements simultaneously and we have obtained the allowed range for the coupling of the scalar neutrino.

\section*{Acknowlegements}

This work was supported by project "Particle Cosmology".

\end{document}